\documentclass[12pt]{article}

\usepackage{epsf}
\usepackage[dvips]{graphicx}
\usepackage{amssymb}

\begin{document}

\begin{center}
{\bfseries POSSIBLE DIFFERENCE BETWEEN MULTIPLICITY DISTRIBUTIONS
AND INCLUSIVE SPECTRA OF SECONDARY HADRONS IN PROTON-PROTON AND
PROTON-ANTIPROTON COLLISIONS AT ENERGY $\sqrt{s}=900$~GeV}

\vskip 5mm

V.A. Abramovsky$^{\dag}$, N.V. Radchenko$^{\ddag}$

\vskip 5mm

{\small {\it Novgorod State University }
\\
$\dag$ {\it E-mail: Victor.Abramovsky@novsu.ru }
\\
$\ddag$ {\it E-mail: nvrad@mail.ru }}
\end{center}

\vskip 5mm

\begin{center}
\begin{minipage}{150mm}
\centerline{\bf Abstract} We consider QCD based model of hadrons
interaction in which gluons density in wave function of initial
state is low in rapidity space and real hadrons are produced by
decay of color field strings. Hadrons production processes in $pp$
and $p\bar{p}$ interactions differ on principle. There are three
types of inelastic processes in $p\bar{p}$ collision. The first
type is production of secondary hadrons shower from decay of gluon
string. The second type is shower produced from decay of two quark
strings and the third one -- from decay of three quark strings. At
the same time there are only two types of inelastic processes for
$pp$ collision, they are shower from gluon string and shower from
two quark strings. Therefore multiplicity distributions and
inclusive spectra of secondary hadrons are different in $pp$ and
$p\bar{p}$ interactions, and this difference may be observed at
energy $\sqrt{s}=900$~GeV.
\end{minipage}
\end{center}

\vskip 10mm

\section{Introduction}
The measurements of the properties of proton-proton collisions at
Large Hadron Collider (LHC) at a center-of-mass energy
$\sqrt{s}=900$~GeV may lead to discovery of ``new physics'' in
soft interactions at high energies. There are sufficiently precise
data for proton-antiproton collisions at the same energy including
charged particle multiplicity distributions $P_n$ and
pseudorapidity distributions ${\rm d}N_{ch}/{\rm
d}\eta$~\cite{bib1, bib2}. The comparison of these values with
forthcoming measurements of LHC will give the opportunity to find
out whether multiple characteristics are the same in $pp$ and
$p\bar{p}$ interactions.

The data obtained by ALICE coll. on $23^{\rm rd}$ November 2009 at
$\sqrt{s}=900$~GeV~\cite{bib3} do not allow to make definite
conclusions because of large uncertainties.

In this article we argue that there are differences in multiple
production in $pp$ and $p\bar{p}$ interactions. These differences
may be observed at $\sqrt{s}=900$~GeV. We give predictions for
absolute value of inclusive cross section ${\rm
d}\sigma^{incl}/{\rm d}\eta$ in Fig.~1 and absolute value of
inclusive cross section ${\rm d}\sigma^{incl}_n/{\rm d}\eta$ for
interval of charged multiplicity $62\leqslant n \leqslant 70$ in
Fig.~2 (all cross sections are in millibarns). Detailed discussion
of these results is in the following sections.

This article is organized as follows: Section 2 describes the
details of our model; elementary inelastic subprocesses are
described in Section 3; Section 4 is dedicated to inclusive
pseudorapidity distributions and discussion of results.

\begin{figure}[!h]
\centerline{
\includegraphics[scale=0.7]{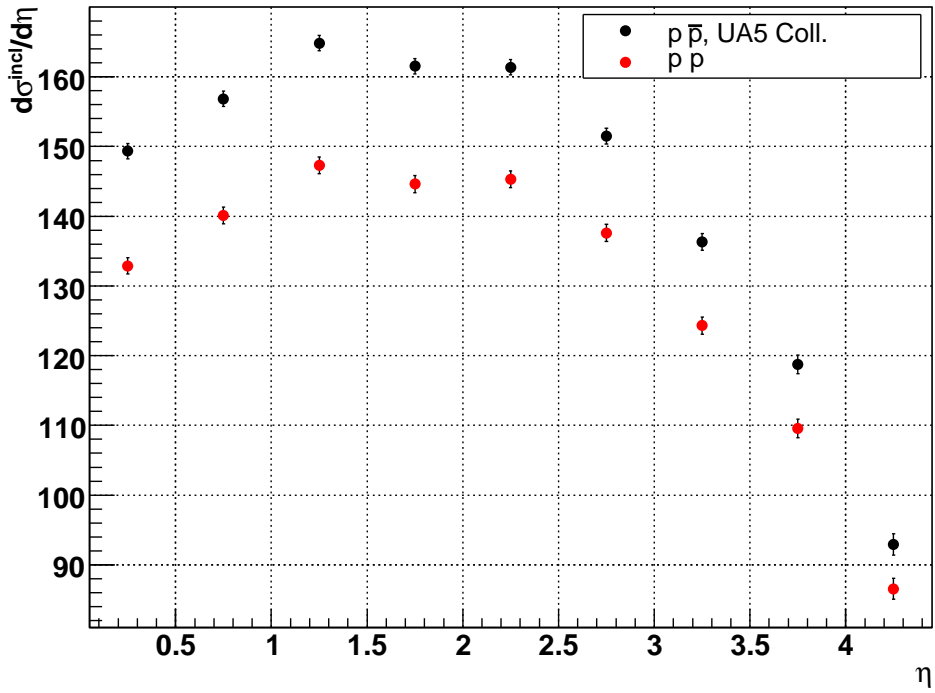}}
\caption{Absolute value of inclusive cross section at
$\sqrt{s}=900$~GeV}
\end{figure}

\begin{figure}[!h]
\centerline{
\includegraphics[scale=0.7]{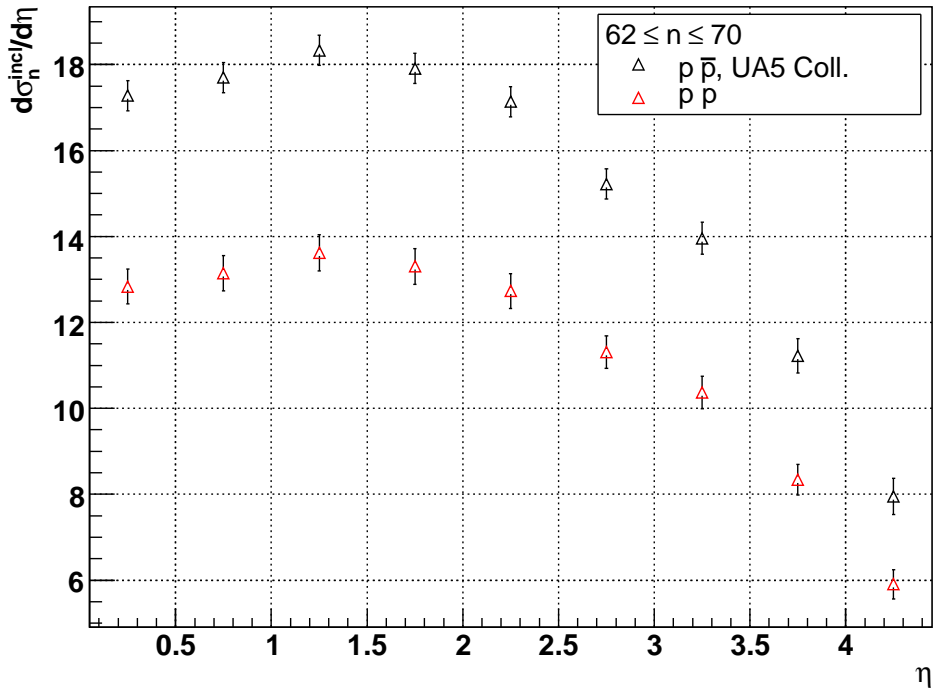}}
\caption{Absolute value of inclusive cross section for charged
multiplicity interval $62\leqslant n \leqslant 70$ at
$\sqrt{s}=900$~GeV}
\end{figure}

\section{Low Constituents Number Model}
It is generally agreed that the same elementary subprocesses
contribute to hadrons production in $pp$ and $p\bar{p}$
interactions at high energies, they are ``pomeron
showers''~\cite{bib4}. These showers correspond to cuts of
different numbers of pomerons, their relations are defined by
Abramovsky-Gribov-Kancheli theorem~\cite{bib5}. Contributions of
non vacuum reggeons die out at high energies. Therefore it is
considered that multiple production characteristics such as
charged particle multiplicity distributions and inclusive spectra
are the same in $pp$ and $p\bar{p}$ collisions. This idea is used
in number of papers, for example~\cite{bib6, bib7}.

We emphasize that inelastic processes  that contribute to $pp$ and
$p\bar{p}$ interactions are different and have nothing in common
with picture of multiple pomeron showers (now they are called
multiple parton interactions). The difference is connected to the
fact that two ``elementary'' inelastic subprocesses contribute to
inelastic production in $pp$ scattering and three ``elementary''
inelastic subprocesses -- in $p\bar{p}$ scattering~\cite{bib8} --
\cite{bib12}.

We explain physics of this phenomenon on the base of Low
Constituents Number Model (LCNM)~\cite{bib13}. It is QCD based
model of hadrons interaction in which gluons density in wave
function of initial state is low in rapidity space and real
hadrons are produced by decay of color field strings. The main
features of this model are the following.

\begin{itemize}
\item Sizes $R$ of hadrons consisting of light quarks are large,
value of coupling constant $\alpha_s(R)$ is large and it is in
region of strong coupling. Therefore there are strong color fields
of non perturbative nature inside hadron. Fragmentation region of
structure function of hadron is filled only with valence quarks,
because transverse gluons have sense only in region of weak
coupling, where the value of coupling constant $\alpha_s(r_g)$ is
small; here $r_g$ -- characteristic sizes for which transverse
real gluons do not overlap with light quarks. These transverse
gluons occur in central region of structure function of fast
hadron as ``bremsstrahlung'' gluons. Since value of
$\alpha_s(r_g)$ is small then number of bremsstrahlung transverse
gluons is small, i.e. their spectrum is sparse.

\item Initial state of hadrons at high energies corresponds to
thin disks with thickness of $1/\sqrt{s}$ with strong color fields
concentrated inside them. When disks get over each other
instantaneous Coulomb exchange takes place and color charge
exchange occurs. Then these thin disks move apart and color
strings stretch between them\footnote{As color field string we
define tube of color field with transverse size much less than
linear size.}.
\end{itemize}

\section{Elementary inelastic subprocesses in $pp$ and $p\bar{p}$ interactions}

Based on these assumptions we supposed in~\cite{bib8, bib9} that
interaction between hadrons is carried out by color exchange of
only one gluon, and there are only one and two additional
bremsstrahlung transverse  gluons in initial state of colliding
hadrons beside valence quarks.

We formalize our treating by introducing diagrams for elementary
subprocesses for $p\bar{p}$ and $pp$ collisions, Fig.~3, 4. Solid
lines correspond to quarks and antiquarks, wavy lines to gluons,
spirals to strings. Interaction in final state is marked by dotted
block.

The diagram in Fig.~3a describes process of $p\bar{p}$ interaction
in case when only valence quarks and antiquarks are in initial
state. This diagram gives constant part of $p\bar{p}$ total cross
section. Interaction occurs as result of gluon exchange between
colorless states. Thereat colorless states (proton and antiproton)
gain octet color charges and move apart. Color field string is
produced between them and when length of string becomes large it
decays into secondary hadrons.

Process of $p\bar{p}$ interaction in case when there are valence
quarks and one additional bremsstra-hlung transverse  gluon in
initial state is shown in diagram in Fig.~3b. One of quark strings
in final state absorbs additional gluon and changes color
charge\footnote{Dynamics of color leading to change of color
charge of quark string will be published later.}. Contribution
from diagram in Fig.~3b to total cross section increases with
energy proportionally to $\ln s$.

Processes of $p\bar{p}$ interaction in case when there are valence
quarks and two additional bremsstrahlung transverse  gluons in
initial state are shown in diagrams in Fig.~3c and 3d. In diagram
in Fig.~3c both gluons are absorbed by the same quark string. It
can be argued~\cite{bib9, bib12} that in such case there are two
quark strings in final state. In diagram in Fig.~3d gluons are
absorbed by different strings. From the same reasons~\cite{bib9,
bib12} there are three quark strings in final state. Contributions
from diagrams in Fig.~3c and 3d to total cross section increase
with energy proportionally to $\ln^2 s$.

\begin{figure}[!h]
\centerline{
\includegraphics[scale=0.7]{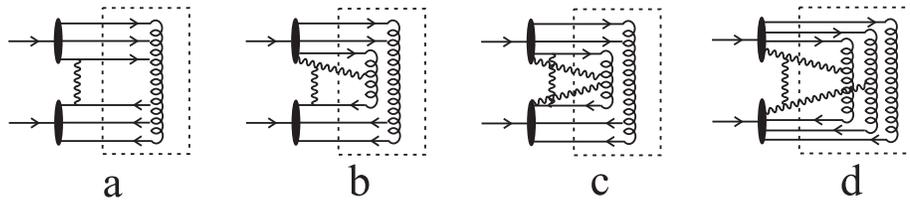}}
\caption{Types of inelastic subprocesses for $p\bar{p}$}
\end{figure}

Elementary processes of $pp$ interaction we describe by diagrams
in Fig.~4.

Process of $pp$ interaction in case when only valence quarks  are
in initial state is shown in Fig.~4a. This diagram gives constant
part of $pp$  total cross section, its contribution completely
coincides with contribution from the corresponding diagram of
$p\bar{p}$ in Fig.~3a.

Diagrams in Fig.~4b and 4c describe hadrons production process in
two quark strings for one and two additional gluons
correspondingly. Gluons are absorbed by one and two color strings
which change color charge. Contribution from diagram in Fig.~4b to
total cross section is proportional to $\ln s$ and coincides with
contribution from diagram in Fig.~3b. Contribution from diagram in
Fig.~4c to total cross section is proportional to $\ln^2 s$.

\begin{figure}[!h]
\centerline{
\includegraphics[scale=0.7]{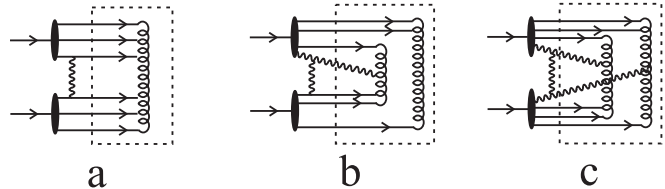}}
\caption{Types of inelastic subprocesses for $pp$}
\end{figure}

In this way in $pp$ interaction quark strings arise between quark
of one proton and diquark of another one therefore state with more
than two quark strings is impossible. At the same time in
$p\bar{p}$ interaction quark strings can be produced between every
quark of  proton and antiquark of antiproton. So in this case
state with three quark strings is possible.

This effect defines difference in multiple characteristics of $pp$
and $p\bar{p}$ interactions.

\section{Inclusive pseudorapidity distributions in $pp$ and $p\bar{p}$ interactions}

In every individual event $n$ charged and $m$ neutral particles
are produced. We do not distinguish sign of charge so we consider
all charged particles as identical and all neutral particles as
identical. Topological cross section of production of $n$ charged
and $m$ neutral particles is defined as
\begin{equation}\label{1}
\sigma_{n+m}=\frac{1}{n!m!}\int{\rm
d}\tau_{n+m}\Bigl|A_{2\rightarrow
n+m}\left(\vec{p}_1,\ldots,\vec{p}_n,\vec{q}_{n+1},\ldots,\vec{q}_{n+m}\right)\Bigr|^2,
\end{equation}
where $A_{2\rightarrow
n+m}\left(\vec{p}_1,\ldots,\vec{p}_n,\vec{q}_{n+1},\ldots,\vec{q}_{n+m}\right)$
-- amplitude of production of $n$ charged and $m$ neutral
particles with corresponding momenta $\vec{p}_i$, $\vec{q}_j$;
${\rm d}\tau_{n+m}$ -- corresponding phase volume. Topological
cross section of production of $n$ charged particles is defined as
\begin{equation}\label{2}
\sigma_{n}=\sum_{m=0}^\infty\sigma_{n+m}=\frac{1}{n!}\sum_{m=0}^\infty\frac{1}{m!}\int{\rm
d}\tau_{n+m}\Bigl|A_{2\rightarrow
n+m}\left(\vec{p}_1,\ldots,\vec{p}_n,\vec{q}_{n+1},\ldots,\vec{q}_{n+m}\right)\Bigr|^2.
\end{equation}
Invariant inclusive cross section of production of one charged
particle in event with $n$ charged particles can be defined as
following:
\begin{equation}\label{3}
(2\pi)^32E_1\frac{{\rm d}^3\sigma_n^{incl}}{{\rm
d}^3p_1}=\frac{1}{(n-1)!}\sum_{m=0}^\infty\frac{1}{m!}\int{\rm
d}\tau_{n-1+m}\Bigl|A_{2\rightarrow
n+m}\left(\vec{p}_1;\vec{p}_2,\ldots,\vec{p}_n,\vec{q}_{n+1},\ldots,\vec{q}_{n+m}\right)\Bigr|^2,
\end{equation}
where integration in ${\rm d}\tau_{n-1+m}$ is performed by all
momenta starting from $\vec{p}_2$, \ldots.

Invariant inclusive cross section can be written as:
\begin{equation}\label{4}
(2\pi)^32E_1\frac{{\rm d}^3\sigma^{incl}}{{\rm
d}^3p_1}=\sum_{n=1}^\infty(2\pi)^32E_1\frac{{\rm
d}^3\sigma_n^{incl}}{{\rm d}^3p_1}.
\end{equation}

Inclusive cross section is normalized by mean multiplicity of
corresponding cross section of inelastic process, here we use non
single diffraction cross section $\sigma^{nsd}$.
\begin{equation}\label{5}
\int{\rm d}^3p_1\frac{{\rm d}^3\sigma^{incl}}{{\rm
d}^3p_1}=\langle n \rangle\,\sigma^{nsd}
\end{equation}
At the same time cross section ${\rm d}^3\sigma^{incl}_n/{\rm
d}^3p_1$ is normalized by the following relation:
\begin{equation}\label{6}
\int{\rm d}^3p_1\frac{{\rm d}^3\sigma^{incl}_n}{{\rm d}^3p_1}= n
\,\sigma_n.
\end{equation}
where $n$ -- number of charged particles in event and $\sigma_n$
-- corresponding topological cross section defined by~(\ref{2}).

We can obtain expressions for inclusive cross sections ${\rm
d}\sigma^{incl}/{\rm d}\eta$ for pseudorapidity $\eta$  (or ${\rm
d}\sigma^{incl}/{\rm d}y$ for rapidity $y$ ) by using integral
of~(\ref{3}) and~(\ref{4}) of transverse components of momentum
$\vec{p}_1$.

Normalization of these cross sections is obvious.
\begin{equation}\label{7}
\int{\rm d}\eta\frac{{\rm d}\sigma^{incl}}{{\rm d}\eta}=\langle n
\rangle\,\sigma^{nsd}
\end{equation}
\begin{equation}\label{8}
\int{\rm d}\eta\frac{{\rm d}\sigma^{incl}_n}{{\rm d}\eta}= n
\,\sigma_n
\end{equation}
We can rewrite~(\ref{8}) as
\begin{equation}\label{9}
\frac{1}{\sigma^{nsd}}\int{\rm d}\eta\frac{{\rm
d}\sigma^{incl}_n}{{\rm d}\eta}= n
\,\frac{\sigma_n}{\sigma^{nsd}}=n\,P_n.
\end{equation}

We think that inclusive cross sections ${\rm d}\sigma^{incl}/{\rm
d}\eta$ (${\rm d}\sigma^{incl}/{\rm d}y$) are the most
informative. Unfortunately, we did not find such experimental data
for $pp$ and $p\bar{p}$ interactions. However, UA5 Collaboration
gave data on inclusive cross sections in nine bins depending on
number of charged particles ($2\leqslant n\leqslant 10$,
$12\leqslant n \leqslant20$, \ldots, $n\geqslant 82$).

We define the following notations:
\begin{equation}\label{10}
\sigma^{(1)}=\sum_{n=2}^{10}\sigma_n,\;\;\;\;\sigma^{(2)}=\sum_{n=12}^{20}\sigma_n,\;\;\ldots\;\;\sigma^{(9)}=\sum_{n=82}^\infty\sigma_n,
\end{equation}
\begin{equation}\label{11}
\sum_{i=1}^9\sigma^{(i)}=\sigma^{nsd}.
\end{equation}

Also we define
\begin{equation}\label{12}
\frac{{\rm d}\sigma^{(1)incl}}{{\rm
d}\eta}=\sum_{n=2}^{10}\frac{{\rm d}\sigma^{incl}_n}{{\rm
d}\eta},\;\;\ldots\;\;\frac{{\rm d}\sigma^{(9)incl}}{{\rm
d}\eta}=\sum_{n=82}^{\infty}\frac{{\rm d}\sigma^{incl}_n}{{\rm
d}\eta},
\end{equation}
\begin{equation}\label{13}
\sum_{i=1}^9\frac{{\rm d}\sigma^{(i)incl}}{{\rm d}\eta}=\frac{{\rm
d}\sigma^{incl}}{{\rm d}\eta}.
\end{equation}

Data of UA5 Collaboration are given in format
\begin{equation}\label{14}
\frac{1}{\sigma^{(i)}}\frac{{\rm d}\sigma^{(i)incl}}{{\rm d}\eta}.
\end{equation}

We integrated the expression
$\displaystyle\frac{1}{\sigma^{nsd}}\frac{{\rm
d}\sigma^{(i)incl}}{{\rm d}\eta}$ over pseudorapidity space
using~(\ref{8}) and obtained
\begin{equation}\label{15}
\frac{1}{\sigma^{nsd}}\int{\rm d}\eta\frac{{\rm
d}\sigma^{(i)incl}}{{\rm
d}\eta}=\frac{1}{\sigma^{nsd}}\sum_{n\;{\rm in\; bin}}\int{\rm
d}\eta\frac{{\rm d}\sigma^{incl}_n}{{\rm d}\eta}=\sum_{n\;{\rm
in\; bin}}n\,\frac{\sigma_n}{\sigma^{nsd}}=\sum_{n\;{\rm in\;
bin}}n\,P_n=\bar{n}^{(i)},
\end{equation}
where $\bar{n}^{(i)}$ is defined as
$$
\bar{n}^{(1)}=\sum_{n=2}^{10}n\,P_n,\;\;\;\;\bar{n}^{(2)}=\sum_{n=12}^{20}n\,P_n,\;\;\ldots\;\;\bar{n}^{(9)}=\sum_{n=82}^\infty
n\,P_n.
$$

Non single diffraction cross sections are the same for $pp$ and
$p\bar{p}$ interactions because of Pomeranchuk theorem. But shapes
of multiplicity distribution curves are different for $pp$ and
$p\bar{p}$ since underlying elementary subprocesses are different
(see Section 3 and Fig.~5). Therefore values of $\bar{n}^{(i)}$
are different for $pp$ and $p\bar{p}$ collisions, we denote them
$\bar{n}^{(i)}_{pp}$ and $\bar{n}^{(i)}_{p\bar{p}}$
correspondingly.

From relation~(\ref{15}) it follows
\begin{equation}\label{16}
\int{\rm d}\eta\frac{{\rm d}\sigma^{(i)incl}_{pp}}{{\rm
d}\eta}\Bigg/\int{\rm d}\eta\frac{{\rm
d}\sigma^{(i)incl}_{p\bar{p}}}{{\rm
d}\eta}=\frac{\bar{n}^{(i)}_{pp}}{\bar{n}^{(i)}_{p\bar{p}}}.
\end{equation}

The expression $\bar{n}^{(i)}_{pp}/\bar{n}^{(i)}_{p\bar{p}}$ does
not depend on pseudorapidity. Besides, number of charged particles
in each bin can be taken arbitrary but relation~(\ref{16}) is kept
strictly. Therefore we have for inclusive cross sections
\begin{equation}\label{17}
\frac{{\rm d}\sigma^{(i)incl}_{pp}}{{\rm d}\eta}\Bigg/\frac{{\rm
d}\sigma^{(i)incl}_{p\bar{p}}}{{\rm
d}\eta}=\frac{\bar{n}^{(i)}_{pp}}{\bar{n}^{(i)}_{p\bar{p}}}.
\end{equation}
From here
\begin{equation}\label{18}
\frac{{\rm d}\sigma^{(i)incl}_{pp}}{{\rm
d}\eta}=\frac{\bar{n}^{(i)}_{pp}}{\bar{n}^{(i)}_{p\bar{p}}}\,\frac{{\rm
d}\sigma^{(i)incl}_{p\bar{p}}}{{\rm d}\eta}.
\end{equation}

From LCNM and fitting of multiplicity distributions for $pp$ and
$p\bar{p}$ at different energies~\cite{bib9, bib11} we obtained
prediction for multiplicity distribution in $pp$ scattering at
$\sqrt{s}=900$~GeV, Fig.~5.

\begin{figure}[!h]
\centerline{
\includegraphics[scale=0.7]{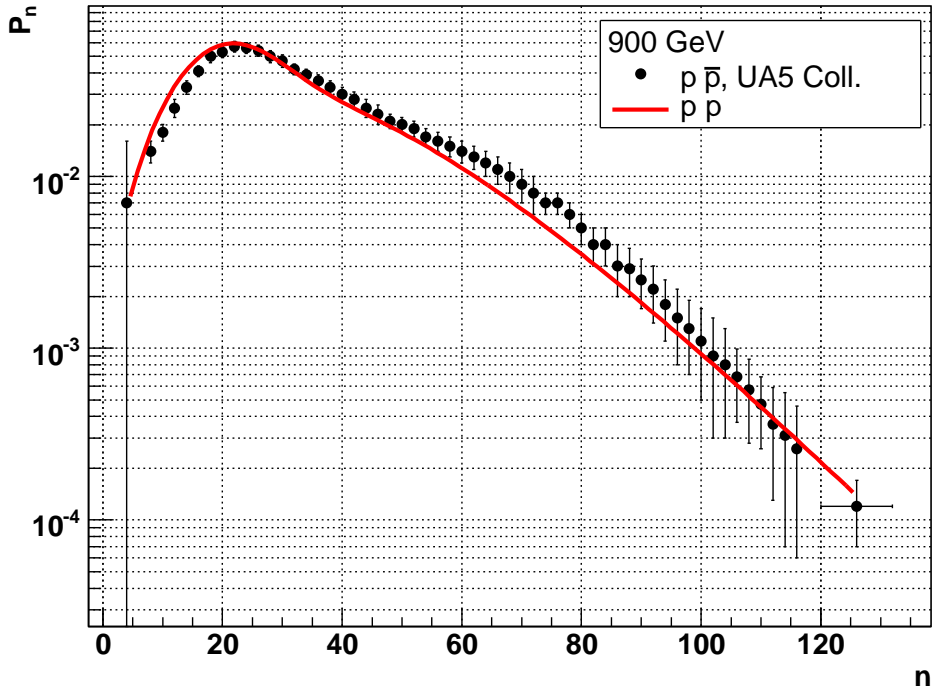}}
\caption{Multiplicity distribution for $p\bar{p}$ at
$\sqrt{s}=900$~GeV (points) and prediction for $pp$ at the same
energy in LCNM (red line)}
\end{figure}

\begin{table}[!h]
\caption{Coefficient for inclusive pseudorapidity distributions}
\begin{tabular}{|c|c|c|c|c|c|}
\hline &$2\leqslant n\leqslant10$&$12\leqslant
n\leqslant20$&$22\leqslant n\leqslant30$&$32\leqslant
n\leqslant40$&$42\leqslant n\leqslant50$\\\hline
$\bar{n}^{(i)}_{pp}/\bar{n}^{(i)}_{p\bar{p}}$&$1.31\pm0.01$&$1.16\pm0.01$&$1.01\pm0.01$&$0.92\pm0.01$&$0.91\pm0.01$\\\hline
&$52\leqslant n\leqslant60$&$62\leqslant
n\leqslant70$&$72\leqslant n\leqslant80$&$n\geqslant82$&\\\hline
$\bar{n}^{(i)}_{pp}/\bar{n}^{(i)}_{p\bar{p}}$&$0.85\pm0.01$&$0.74\pm0.02$&$0.69\pm0.01$&$0.79\pm0.02$&\\\hline
\end{tabular}
\end{table}

We calculated the values of coefficients
$\bar{n}^{(i)}_{pp}/\bar{n}^{(i)}_{p\bar{p}}$ for nine bins of
multiplicity to estimate the difference in inclusive cross
sections for $pp$ and $p\bar{p}$ with~(\ref{18}), Table~1. Values
of probabilities $P_n$ for $pp$ we took from our prediction in
case when 75\% of two gluons give three quark strings in
$p\bar{p}$, values of $P_n$ for $p\bar{p}$ we took from UA5
experiment~\cite{bib1}. Inclusive pseudorapidity distributions for
all bins of charged multiplicity are shown in Fig.~6-13 and
Fig.~2. From~(\ref{13}) we obtained the absolute value of
inclusive pseudorapidity distribution, shown in Fig.~1.

Quite obviously that inclusive distribution give less visible
difference than inclusive distributions in different bins. This is
due to fact that multiplicity distribution for $pp$ has higher
values before peak and in peak, but lower values in tail of
distribution than for $p\bar{p}$. So difference is compensated in
sum.

In conclusion we want to stress that difference between $pp$ and
$p\bar{p}$ interactions may be observed even at energy
$\sqrt{s}=900$~GeV, especially in inclusive distributions in
different bins of charged multiplicity.

\begin{figure}[!h]
\centerline{
\includegraphics[scale=0.6]{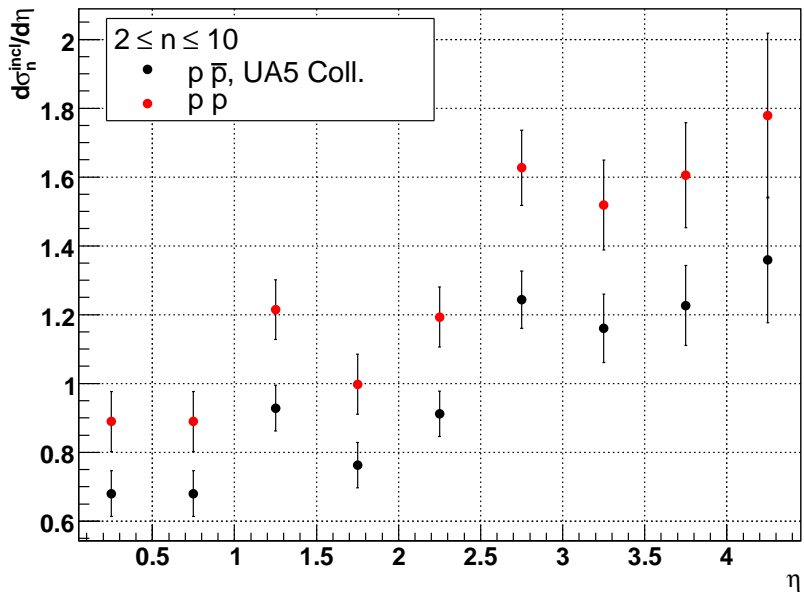}}
\caption{Inclusive pseudorapidity distributions for $2 \leqslant
n\leqslant 10$}
\end{figure}
\begin{figure}[!h]
\centerline{
\includegraphics[scale=0.6]{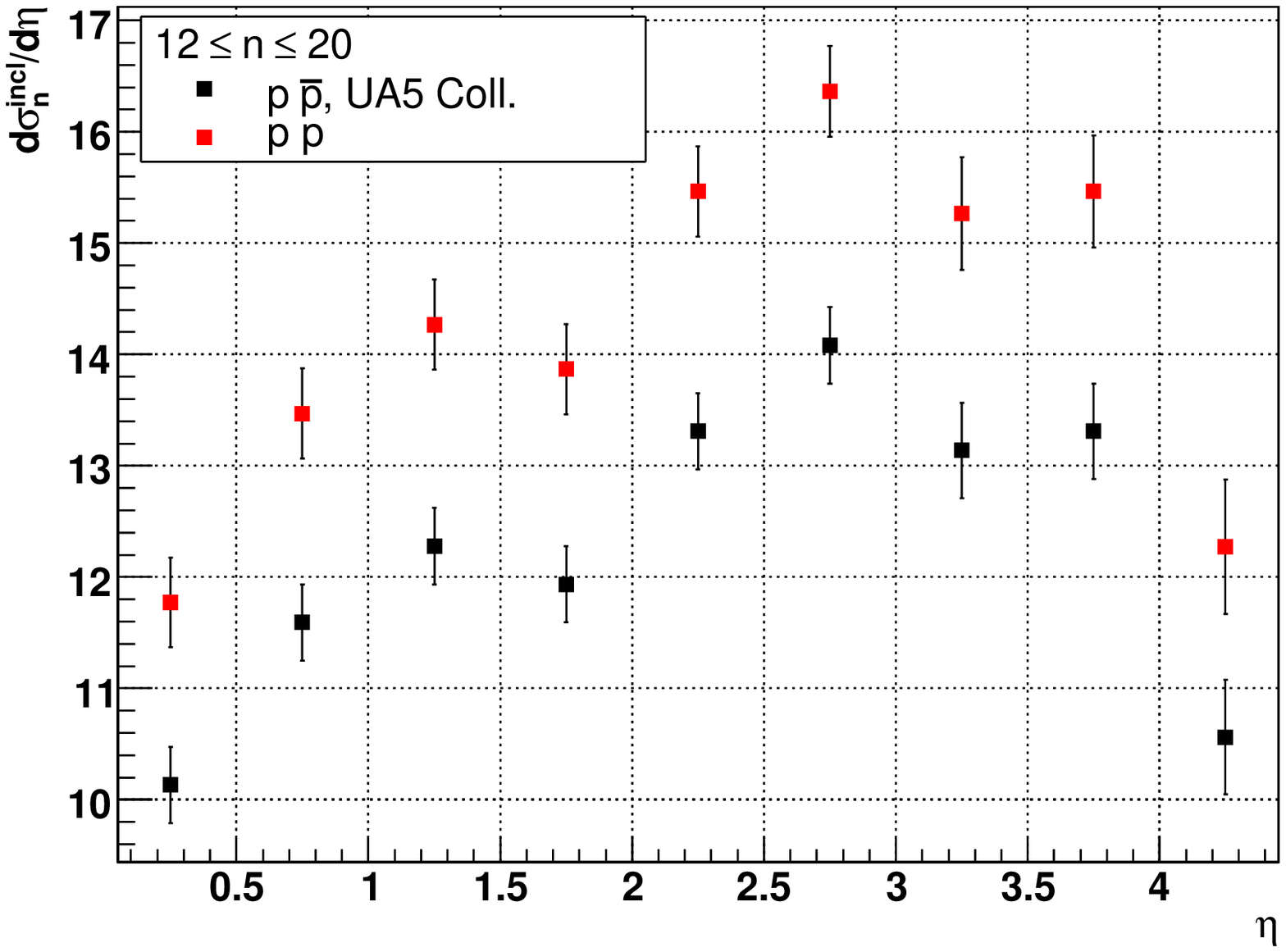}}
\caption{Inclusive pseudorapidity distributions for $12 \leqslant
n\leqslant 20$}
\end{figure}
\begin{figure}[!h]
\centerline{
\includegraphics[scale=0.6]{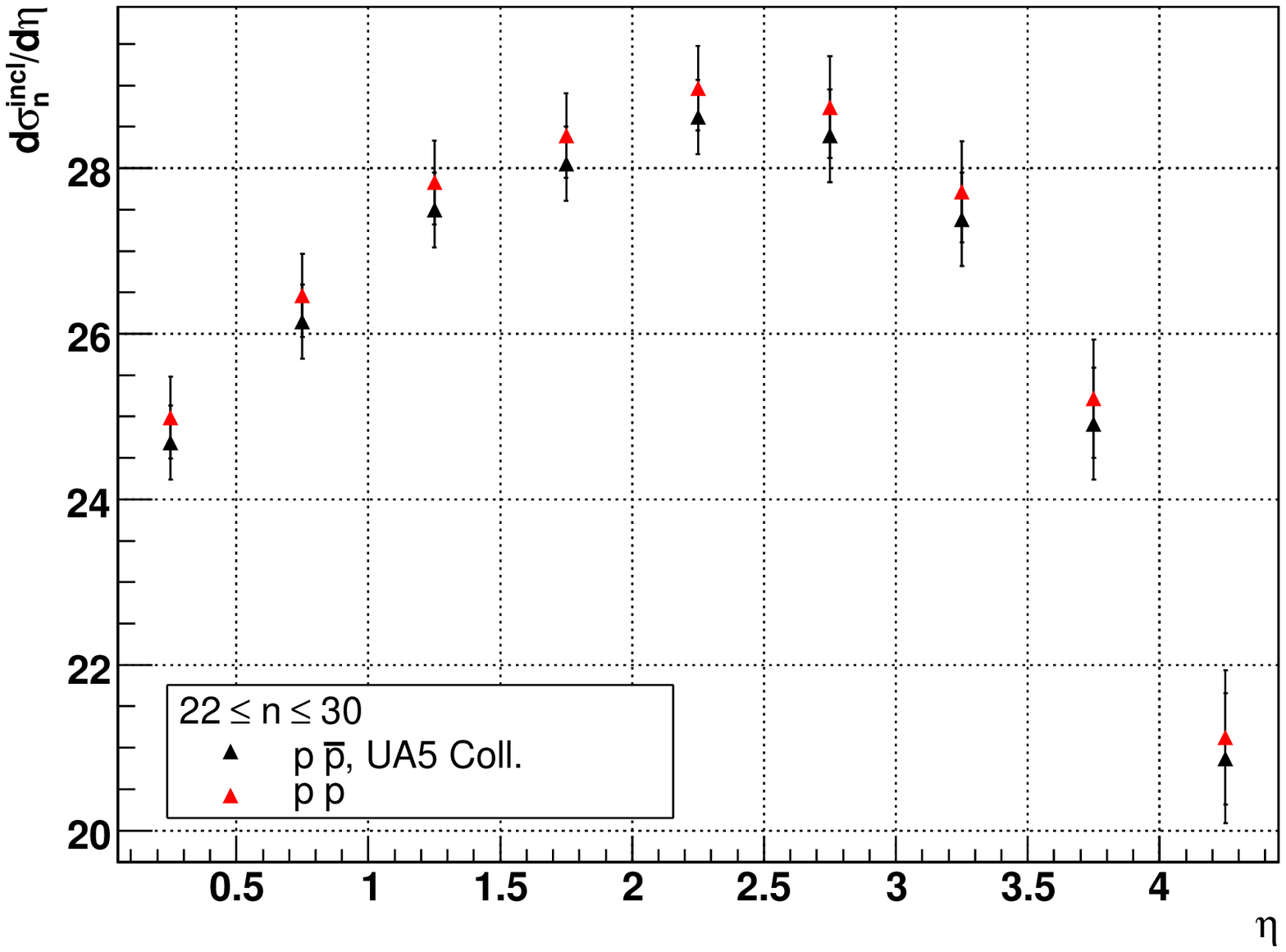}}
\caption{Inclusive pseudorapidity distributions for $22 \leqslant
n\leqslant 30$}
\end{figure}
\begin{figure}[!h]
\centerline{
\includegraphics[scale=0.6]{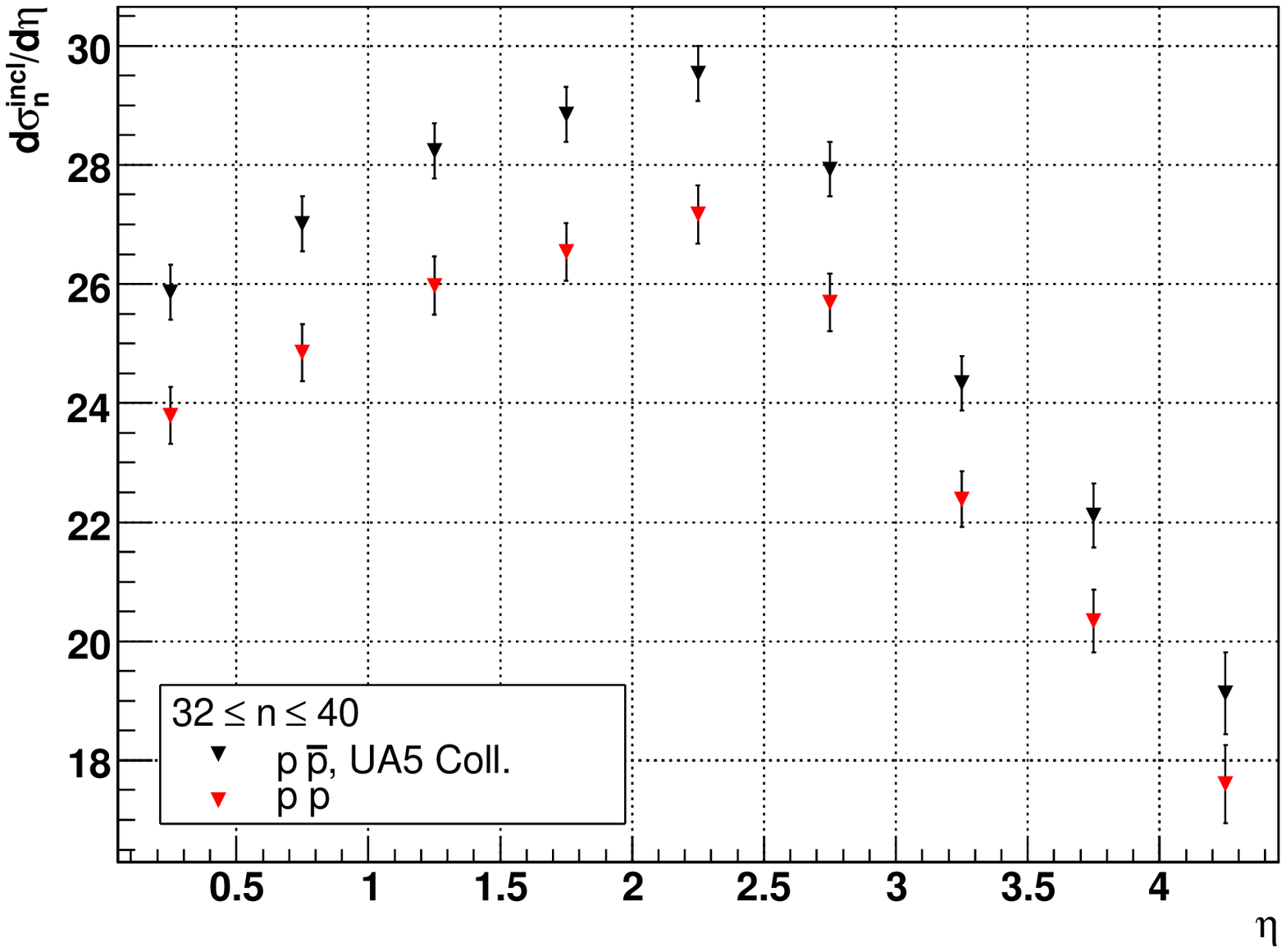}}
\caption{Inclusive pseudorapidity distributions for $32 \leqslant
n\leqslant 40$}
\end{figure}
\begin{figure}[!h]
\centerline{
\includegraphics[scale=0.6]{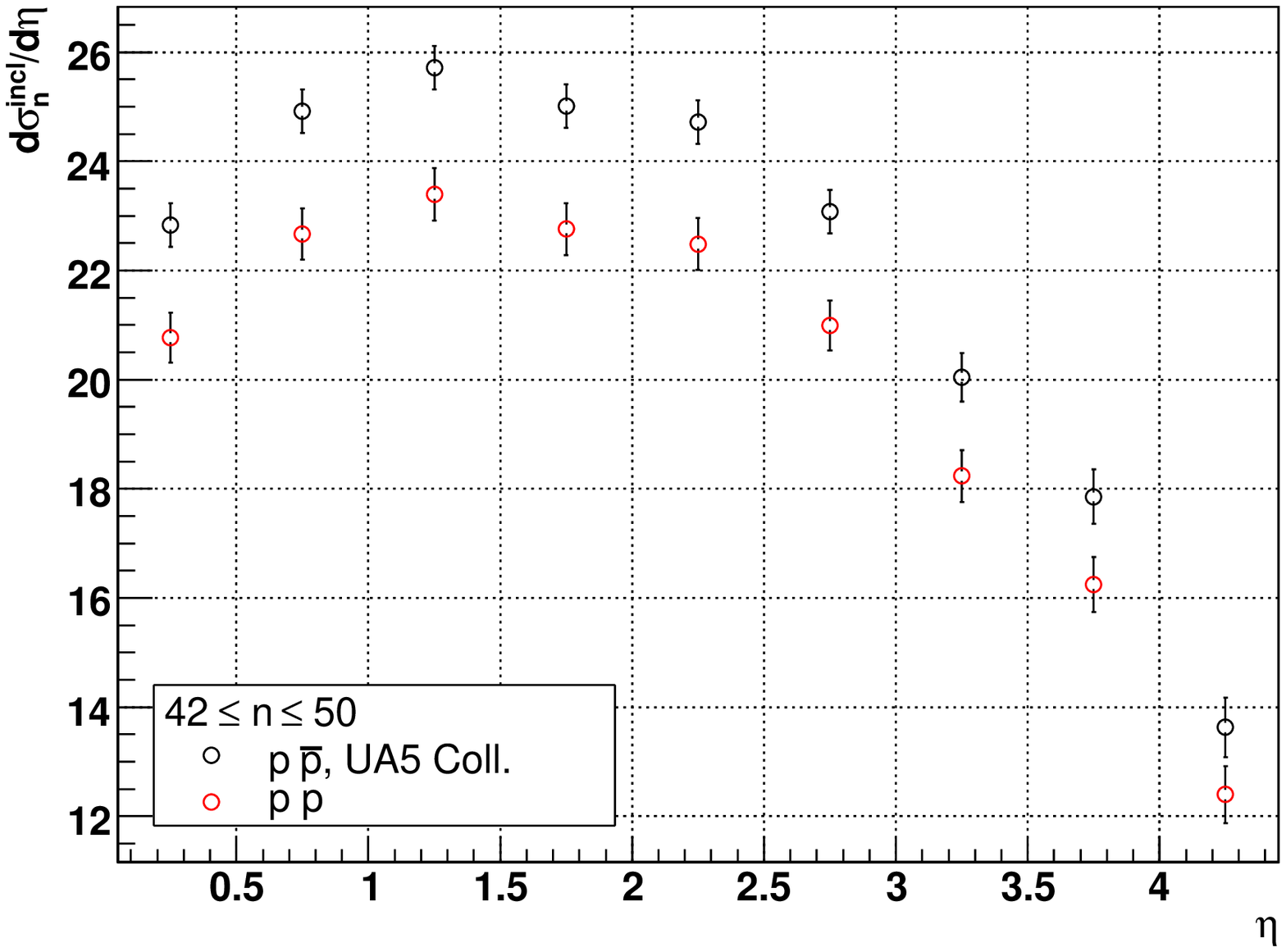}}
\caption{Inclusive pseudorapidity distributions for $42 \leqslant
n\leqslant 50$}
\end{figure}
\begin{figure}[!h]
\centerline{
\includegraphics[scale=0.6]{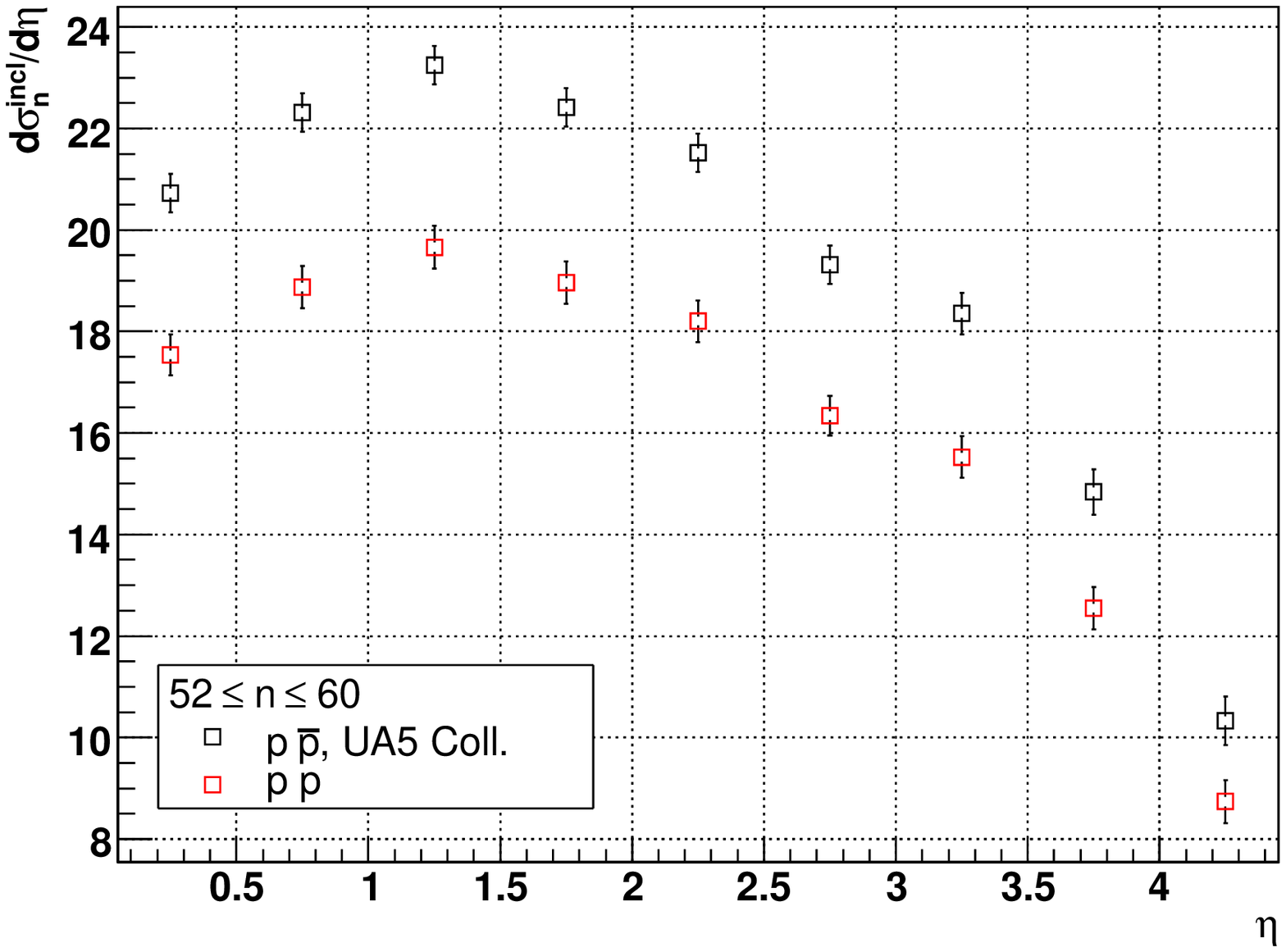}}
\caption{Inclusive pseudorapidity distributions for $52 \leqslant
n\leqslant 60$}
\end{figure}
\begin{figure}[!h]
\centerline{
\includegraphics[scale=0.6]{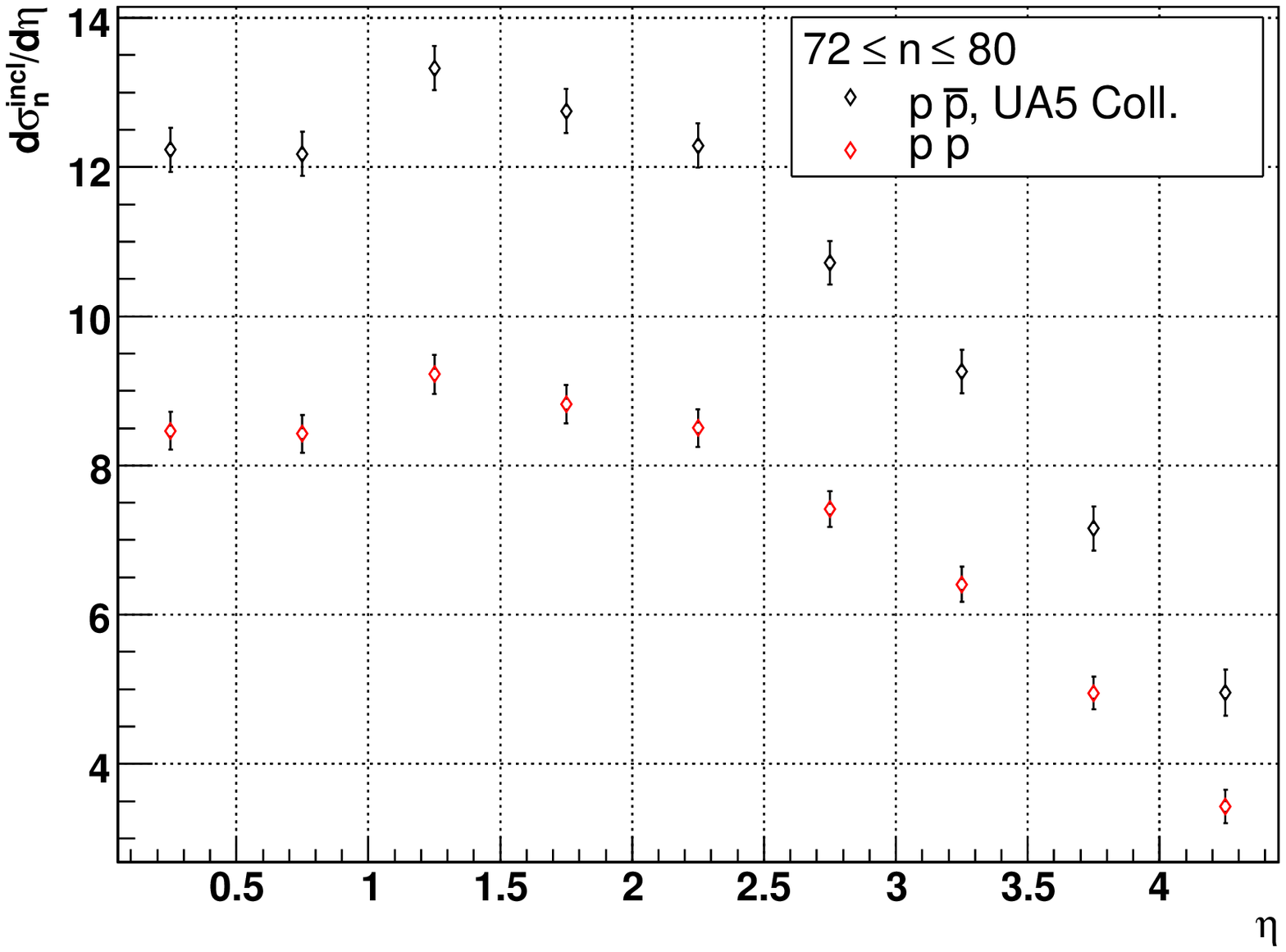}}
\caption{Inclusive pseudorapidity distributions for $72 \leqslant
n\leqslant 80$}
\end{figure}
\clearpage
\begin{figure}[!h]
\centerline{
\includegraphics[scale=0.6]{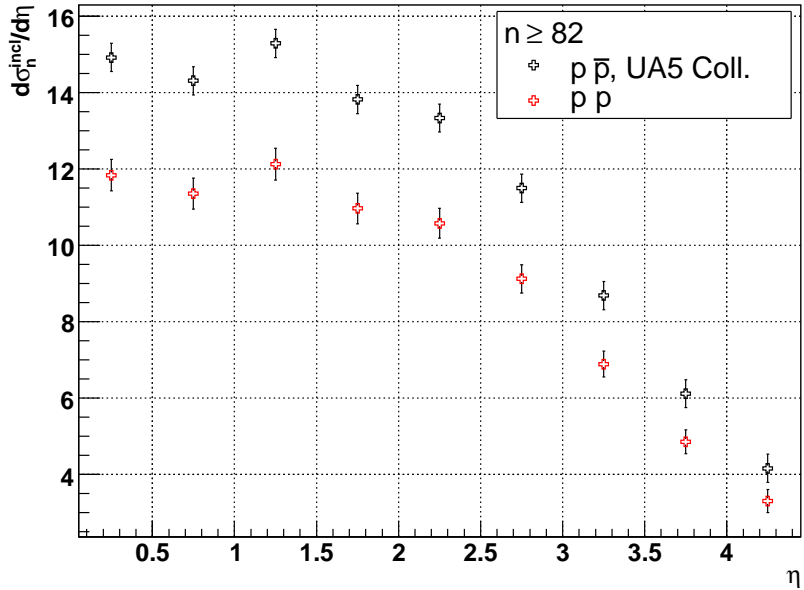}}
\caption{Inclusive pseudorapidity distributions for
$n\geqslant82$}
\end{figure}

\end{document}